\documentclass[A4, 11pt]{article}
\usepackage{booktabs} 

\usepackage[colorlinks=true,linkcolor=blue,citecolor=blue]{hyperref}
\usepackage[utf8]{inputenx}
\usepackage{wrapfig}

\usepackage{pgfplots}
\usepackage{graphicx,wrapfig}
\usepackage{amsmath,amsfonts}
\usepackage{mathrsfs}
\usepackage[left=1in,top=1in,right=1in,bottom=1in]{geometry}
\usepackage{float}
\usepackage{mathtools}
\usepackage{listings}
\usepackage{color}
\usepackage[english]{babel}
\usepackage{amssymb}
\usepackage{amsthm}
\usepackage[ruled,vlined,linesnumbered]{algorithm2e}

\SetAlFnt{\small}
\SetAlCapFnt{\small}
\SetAlCapNameFnt{\small}
\SetAlCapHSkip{0pt}
\IncMargin{-\parindent} 
\usepackage{nicefrac}
\usepackage{natbib}

\usepackage{eurosym}
\usepackage{sidecap}
\usepackage{subcaption}
\usepackage{fancyhdr}
\usepackage{multirow}
\usepackage{enumitem}
\usepackage[skip = 6pt]{parskip}
\usepackage{tagging}
\usepackage{tikz}
\usepackage[T1]{fontenc}
\usepackage{cleveref}
\usepackage{textcomp}

\usepackage{bm}
\usepackage{bbm}

\usepackage{array}

\allowdisplaybreaks

\usepackage{footmisc}
\usepackage[most]{tcolorbox}
\usepackage{tabularx}
\usepackage{eurosym}

\newcommand{\newmacro}[2]{\newcommand{#1}{\debug{#2}}}		

\DeclarePairedDelimiterX{\setdef}[2]{\{}{\}}{#1:#2}

\newmacro{\dd}{\:d}		

\theoremstyle{definition}{

}

\title{\textbf{Algorithmic Shortlisting in Participatory Budgeting}}

\author{\textbf{Juan Ignacio Zambrano}$^{1,2}$, \textbf{Clément Contet}$^{1,3}$, \textbf{Jairo Gudiño-Rosero}$^{1,2,4}$,\\
\textbf{Felipe Garrido-Lucero}$^{1,2}$, \textbf{Umberto Grandi}$^{1,2}$, \textbf{César A. Hidalgo}$^{4,5,6}$\\ [0.8em]
\small\textsuperscript{\rm 1}IRIT, Toulouse, France\\
\small\textsuperscript{\rm 2}Université Toulouse Capitole, Toulouse, France\\
\small\textsuperscript{\rm 3}Université de Toulouse, Toulouse, France\\
\small\textsuperscript{\rm 4}Center for Collective Learning, IAST, Toulouse School of Economics,\\
\small{Université Toulouse Capitole, Toulouse, France}\\
\small\textsuperscript{\rm 5}Center for Collective Learning, CIAS, Corvinus University of Budapest, Budapest, Hungary\\
\small\textsuperscript{\rm 6}AMBS, University of Manchester, Manchester, UK
}

\date{}
        
\begin{document}

\maketitle

\begin{abstract}
\noindent Participatory budgeting is a democratic innovation that allows citizens to propose and vote on public investment projects. To help organizers manage large volumes of submissions, we design and test AI methods for algorithmic shortlisting. Our algorithms predict which projects are likely to be funded using only project features and anonymous historical voting data. We demonstrate the limitations of a naive approach that uses a large language model to rank projects based on past success and propose a vote-based pipeline that enables state-of-the-art LLMs to perform on par with classical machine learning. Our findings indicate that user preferences in participatory budgeting are stable enough to allow algorithmic shortlisting to approximate an initial selection of projects effectively.

\medskip

\noindent \textbf{Code availability.}
The original code, prompts, and datasets used in this study can be accessed at
\url{https://doi.org/10.5281/zenodo.21829180}.
\end{abstract}


\section{Introduction}\label{sec:introduction}
    
    
Cities around the world are increasingly giving citizens a more active role in public decision-making \citep{mitchell1996city,peixoto2016does,helbing2024co} giving rise to a number of \emph{democratic innovations} \citep{smith2009democratic}, ranging from in-person processes such as deliberative mini-publics \citep{farrell2019deliberative} to digital participatory experiments such as petition platforms and participatory budgeting \citep{hagen2018content,wampler2021participatory, li2025scaling}.
    A common factor of these participatory initiatives is the presence of large volumes of textual data, posing significant challenges in terms of analysis and scalability.
    
    In this context, recent advances in large language models (LLMs) have enabled new approaches to democratic citizen participation. 
    Notable examples include the popular deliberative platform Pol.is \citep{small2023opportunities}, with LLMs producing a report of the 3,940 ideas commented and voted by 8,000 residents during a digital large-scale town-hall meeting in Bowling Green, U.S. \citep{BowlingGreen}, or the use of LLMs to summarize contrasting opinions in deliberative experiments aimed at finding consensual statements on policy questions \citep{tessler2024ai,konya2025using}.
    
    
    In this article, we focus on \emph{participatory budgeting} (PB), a widely adopted democratic process by which a portion of a municipal or regional budget is allocated to projects proposed and selected by its residents. PB processes are usually composed of three main phases \citep{wampler2000guide,cabannes2004participatory,shah2007participatory}: a \textit{proposal phase}, where citizens submit project ideas and their costs are estimated by an expert entity; a \textit{consolidation phase}, where proposals are shortlisted after removing infeasible or redundant projects; and a \textit{voting phase}, where citizens select from among the finalists. At the end of the voting phase, votes are aggregated and combined with budget constraints to determine the winning projects using a predefined selection rule.
    
    Researchers in computational social choice have focused on designing and testing rules for the voting phase (see, e.g., \citet{TalmonFaliszewski2019,goel2019knapsack,PierczynskiSP21}). In contrast, little attention has been given to the consolidation phase, with the notable exception of \citet{ReyEndrissHaan2021}. Yet, in most PB campaigns, this phase remains rather obscure and poorly documented, typically involving in-person committees taking decisions without well-specified objectives.\footnote{To get an example of scale, the City of Cambridge, MA, U.S., 
    received over 1300 project proposals and shortlisted 20 of them in a consolidation process lasting months \citep{CambridgeParticipatoryBudgeting}.} We claim that a fully specified and documented algorithmic process for shortlisting could substantially improve transparency while reducing the costs associated with processing proposed projects, e.g., through a reliable and testable pre-screening mechanism.
    
    A typical project proposal in PB is characterized by its name, description, category, geographical location, and associated cost (with both name and description being given by the proposer), 
    constituting a rich corpus of textual data. The use of LLMs for treating project proposals therefore arises as a natural idea. 
    \citet{damle2025llms}, for instance, tested the ability of LLMs to select winning projects from community metadata such as age, education, and location. Their work, however, pushes AI usage to a boundary of legitimacy that might be hard to accept by PB organizers or by the population as, e.g., their tools might be used to run PB elections without asking the citizens to actually vote.
    
    Our objective is different: We aim at building a support-tool for PB organizers trying to minimize the AI intervention and leaving voters with the last say on winning projects. Motivated by the observation that a naive LLM shortlisting mechanism may fail to identify the most popular projects, we focus on estimating the success of project proposals in terms of number of votes, producing a coarse pre-ranking of projects to be used, e.g., when prioritizing those that might pass to feasibility study. 
    Our methodology leverages historical voting data, as PB campaigns are often repeated in each city, and the rich textual and tabular data within project proposals. 
    Importantly, we rely only on publicly available data, designing predictors that comply with principles of data minimization,
    essential in democratic applications to build trust and to encourage participation.
    \smallskip
    
    \noindent\textbf{Contributions}. Our contributions are the following.
    \smallskip
    
    \noindent$\bullet$ We show that a naive approach of prompt-based shortlisting is largely inefficient when tested on six target PB campaigns in three European cities with different linguistic and cultural contexts (Toulouse, France, Wroclaw and Warszawa, Poland) which shared their data  for multiple years on Pabulib.org \citep{FaliszewskiEtAlIJCAI2023}.
    
    \noindent$\bullet$ We combine two classical machine learning approaches (ElasticNet and XGBoost) with textual preprocessing strategies, testing their capability to order projects by number of votes in a PB campaign when trained on the outcomes of the previous campaigns. 
    
    
    \noindent$\bullet$ We design three prompting strategies on off-the-shelf LLMs (GPT-4o mini and Mistral Large 3) provided with progressively richer context from previous PB campaigns, testing their capability to order projects by number of votes in subsequent campaigns.
    
    \noindent$\bullet$ We construct and release datasets covering multi-year PB processes for the cities of Toulouse, France, and Wroclaw, Poland, combining voting data from Pabulib.org with project descriptions publicly available on the organisers' webpages.
    
    \smallskip
    
Our results show that our pipeline for vote-based algorithmic shortlisting can meaningfully predict project success albeit with important limitations. The performance of LLMs is comparable to classical machine learning methods but requires a retrieval-augmented generation (RAG) setup to obtain consistent results, thus requiring computational skills that are typically not common in PB organising bodies. 
    We also show that textual project information can improve predictive performance in some settings, highlighting the potential of augmenting classical voting datasets such as Preflib.org and Pabulib.org with textual descriptions of the alternatives \citep{grandi:hal-05548117}.
    This further highlight the potential of 
    our approach, since our results match and often surpass those obtained using LLM personas based on participants' demographic data \citep{yang2024llm}, demonstrating that historical data from same-city PB processes can effectively replace demographic inputs.
    
    \smallskip
    \noindent\textbf{Related Work}. Citizens’ involvement in proposing and selecting public projects began in Latin America in the 1980s and later spread globally.
    The AI research community has focused on the mechanism design and computational aspects of this process, with PB viewed as a form of multi-winner voting under budget constraints \citep{lackner2023multi}.
    PB has been extensively studied in computational social choice (see the recent survey by \citeauthor{rey2025computationalsocialchoiceindivisible}, \citeyear{rey2025computationalsocialchoiceindivisible}).
    Key contributions include the analysis of proportionality axioms (e.g., \citeauthor{aziz2021proportionally}, \citeyear{aziz2021proportionally}) and the development of a fair and computationally tractable PB rule \citep{PierczynskiSP21}, which has been recently tested in European cities and in behavioral experiments \citep{yang2024designing}.\footnote{See also \url{https://equalshares.net/}}
    \citet{ReyEndrissHaan2021} are the only ones to formally study a PB model including both a consolidation phase and a voting phase, proposing aggregation functions for shortlisting and analyzing incentive-compatibility.
    
    
    \citet{arana2021citizen} and \citet{GrossiDigitalDem2024} argue that a range of computational methods must be mobilized for digital democracy to succeed.
    LLMs are among the most actively explored tools in this domain, from generating consensus statements in deliberative settings \citep{tessler2024ai,konya2025using}, to enhancing participatory platforms \citep{davies2021evaluating,small2023opportunities}, to selecting proposals that meet representation guarantees \citep{fish2024generative}.
    An important approach employs LLMs to create voting personas or digital twins—autonomous agents replicating individual voter behavior—within the broader framework of augmented democracy \citep{HidalgoAugmented,grandi2019agent,garcia2024digital}.
    Key examples include \citeauthor{gudino2024large} (\citeyear{gudino2024large}), who augmented voting data on Brazilian government proposals using fine-tuned LLMs, and \citeauthor{yang2024llm} (\citeyear{yang2024llm}), who simulated a voter population in a PB experiment using LLM personas generated via prompting.
    Constructing LLM personas typically requires demographic data, at odds with our goal of building a PB-support tool that relies solely on publicly available data.

    

      \begin{table*}
      \centering
        \begin{sc}
        \small
          \begin{tabularx}{\textwidth}{l p{2.7cm} X l r c r}
            \toprule
            ID & Name & Description & Category & Cost & Dist. & Votes \\
            \midrule
            201 & Ping-pong tables and volleyball courts & Movement is health. In Wroclaw, there is a high demand for places where one can actively spend time outdoors. ... & Sport & 1 000 000 & 1 & 3 974 \\
            \addlinespace
            694 & "Green bus stops" in the city center & The project proposes the creation of green bus stops at selected locations in downtown Wroclaw by planting climbing plants... & Greenery & 200 000 & 1 & 943 \\
            \bottomrule
          \end{tabularx}
        \end{sc}
    \caption{Example of two projects translated to English from the 2017 Wroclaw PB campaign. Each project is characterized by an ID, a name, a textual description, a category, its cost, the city district, and the number of votes received.}
    \label{tab:project_examples}
    \end{table*}

\section{Preliminaries}
This section is dedicated to explain the context of a participatory budgeting process, the datasets we have used during the article, and to show how LLM-naive-shortlisting might struggle to find the most popular projects.
    
\subsection{Context: Participatory Budgeting} 
    
    A PB campaign starts with the announcement of an available budget by the organising body, followed by the collection of project proposals from the population.
    A shortlisting phase follows, 
    and eligible citizens are then invited to vote on the consolidated list of projects during a pre-defined time window.
    Citizens are typically asked to select up to $k$ projects ($k = 3$ in Toulouse, $k=2$ in Wroclaw, and $k=10$ in Warszawa, the cities analyzed in this paper), either online or on-site.
    Once votes have been collected, a voting rule selects a set of projects to be funded given the budget constraint and the elicited preferences. 
    The greedy rule\footnote{The greedy rule selects projects in order of vote count, each time updating the remaining budget and skipping a project if there is not enough remaining budget to fund it \citep{aziz2020participatory}.} is the most commonly used by municipalities due to its simplicity and explainability, but there are many other ways to select winning projects in PB, such as the method of equal shares \citep{PierczynskiSP21}, quadratic voting \citep{lalley2016quadratic}, or knapsack methods \citep{goel2019knapsack}.
    
    \subsection{Multi-Year Datasets}
    
    Voting data for PB campaigns is often made available by organisers, and a number of datasets have been collected on the public repository Pabulib.org \citep{FaliszewskiEtAlIJCAI2023}. 
    Our objective of developing PB predictors requires multi-year datasets of PB instances in the same city using the same ballot format. Among those available on Pabulib, at least three cities satisfy these requirements: Toulouse 2022 and 2024 in France, Wroclaw 2016, 2017, and 2018 in Poland, and Warszawa 2021, 2022, 2023, and 2024 in Poland. We consolidated datasets for these cities using approval-election instances from Pabulib, which typically include a unique project ID, the project name, category, estimated cost, district or neighborhood, and vote count.
    For Toulouse 2022 and 2024 and Wroclaw 2016 and 2017, we further augmented the data with textual descriptions of each project, collected by web scraping on the webpages of PB organizers. These descriptions are exploited in the last experiment of the article. 
    Examples of projects (with descriptions) can be seen in Table~\ref{tab:project_examples}. More examples, together with a detailed description of the datasets and their sources, are provided in Section~\ref{app:datasets} of the Supplementary Material.

    For the city of Toulouse, our analysis focuses on the PB campaigns held in 2022 and 2024, during which 200 and 183 projects, respectively, were short-listed and submitted for public voting.
    For the city of Wroclaw, our analysis focuses on the PB elections held in 2016, 2017, and 2018, during which 52, 50, and 39 projects, respectively, were short-listed and submitted to public vote.
    It is important to note that the Wroclaw dataset in Pabulib includes some demographic information about voters which we removed to ensure consistency. Finally, for the city of Warszawa, our analysis focuses on the PB campaigns held from 2021 to 2024, during which 106, 129, 138, and 118 projects, respectively, were short-listed and submitted to public vote. Table \ref{tab:participation_summary} provides details of the mentioned campaigns.\footnote{We converted budgets to euros using the 2026 exchange rate of 1 zł = 0.24 €.}
    
    In Toulouse, voters supported on average 3 projects in each ballot, giving a total number of 11,606 votes in 2022 and 21,780 in 2024. In Wroclaw, voters could support up to 2 projects, giving 119,194 total votes in 2016, 111,961 in 2017, and 94,995 in 2018. In Warszawa, voters could support up to 10 projects, giving 958,990 total votes in 2021, 812,340 in 2022, 789,040 in 2023, and 752,550 in 2024. The budgets in Toulouse and Warszawa are 6 to 8 times the budget in Wroclaw. Regarding project's average cost in euros, projects in Wroclaw have on average half the cost of those in Warszawa and have higher costs by a factor of 1.8 than those in Toulouse. 

    \medskip

    \begin{table}[ht]
    \centering
    \small
    \setlength{\tabcolsep}{4pt}
    \begin{tabular}{@{}lccrr@{}}
    \hline
    City & Year & Projects & Voters & Budget \\
    \hline
    Wroclaw, PL  & 2016 & 52  & 67\,103 & 1 M \euro \\
    Wroclaw, PL  & 2017 & 50  & 62\,529 & 900k \euro \\
    Wroclaw, PL  & 2018 & 39  & 53\,801 & 900k \euro \\
    \hline
    Toulouse, FR & 2022 & 200 & 4\,532 & 8.0 M \euro \\
    Toulouse, FR & 2024 & 183 & 7\,260 & 8.0 M \euro \\
    \hline
    Warszawa, PL & 2021 & 106 & 95\,899 & 5.8 M \euro \\
    Warszawa, PL & 2022 & 129 & 81\,234 & 6.6 M \euro \\
    Warszawa, PL & 2023 & 138 & 78\,904 & 7.1 M \euro \\
    Warszawa, PL & 2024 & 118 & 75\,255 & 7.1 M \euro \\
    \hline
    \end{tabular}
    \caption{Summary of the PB campaigns studied in this article.}
    \label{tab:participation_summary}
    \end{table}

    \subsection{Naive Shortlisting}
    \begin{figure*}[ht]
        \centering
        \includegraphics[width=\linewidth]{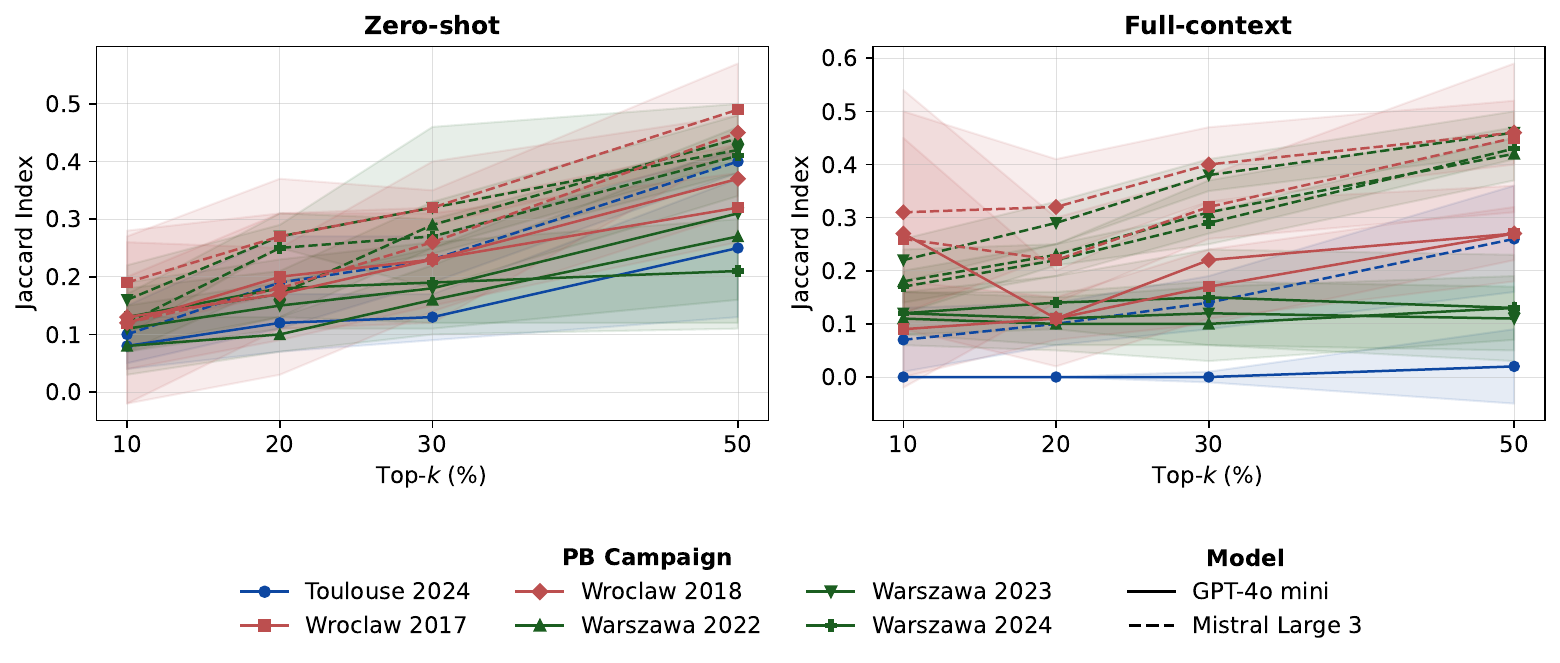}
        \caption{Jaccard index of naive shortlisting for two LLMs, with and without context, over the six PB campaigns. Results are averaged over 20 runs plotted with standard deviation. Performance has high variability and is close to random for all $k$.}
        \label{fig:naive_shortlisting}
    \end{figure*}
    
    Datasets in PB campaigns present an important textual dimension. Additionally, the easy access to the use of LLMs by non-experts motivate us to assess the performance of naively trying to shortlist the projects by simply prompting the task to an LLM. Since the whole list of proposed projects (before the actual shortlisting) is not available, we asked two out-of-the-box LLMs, namely, GPT-4o mini and Mistral Large 3, to select the top $k\%$ of the projects with $k \in \{10,20,30,50\}$, that is, to make a shortlisting of the already shortlisted list, in order to check whether they are capable to find the most popular projects. Moreover, we considered two versions for each LLM, namely, zero-shot, where the LLM receives no context, and full-context, where the LLM receives the list of winners of the previous campaign. Note that, for the sake of comparability, none of the predictions nor context included the projects' descriptions.

    Figure~\ref{fig:naive_shortlisting} shows the Jaccard index (cf. Equation~(\ref{eq:jaccard_index})) obtained by naive shortlisting in six of our instances (the oldest PB campaign of each city is not considered as the ``previous campaign" does not exist). We observe that all methods struggle to identify popular projects. Surprisingly, the performance worsens when presented with context from previous years, in part due to a higher number of LLM hallucinations, showing that a more clever way of exploiting textual information and past voting data is required. See Figure~\ref{fig:shortlisting_prompts} for a detailed explanation of the prompts used as well as any computational specification.
    
    Since a naive use of LLMs for shortlisting fails, in the following section we introduce a new shortlisting methodology based on predicting the vote count of each project.
    
    
    
    

    \section{Vote-Based Shortlisting}\label{sec:metrics}
    
    
    We propose a novel methodology based on predicting the level of public support, given by the vote count, of each project as an intermediate step for PB shortlisting. 
    Given the projects submitted to public voting in a target PB campaign, we first predict the number of votes that each project will receive using either machine learning models or LLM-based predictors with different levels of contextual information. 
    The predicted vote counts are then used to rank projects in descending order of expected public support. 
    We compare this predicted ranking with the actual ranking observed in the corresponding PB campaign. 
    
    
    
    
     
    To test whether our predictions are accurate,
    we compute the \textbf{Kendall’s Tau coefficient} $\tau \in [-1,1]$ between the real and predicted rankings, where $\tau = 1$ denotes perfect agreement and $\tau = -1$ denotes perfect disagreement. 
    To get a more fine-grained analysis of our predictors, we also measure the agreement between the real and predicted rankings using the \textbf{Jaccard index}:
    \begin{align}\label{eq:jaccard_index}
    J(\mathrm{t\widehat{o}p}_k,\mathrm{top}_k) := \frac{|\mathrm{t\widehat{o}p}_k \cap \mathrm{top}_k|}{|\mathrm{t\widehat{o}p}_k \cup \mathrm{top}_k|},
    \end{align}
    where $\mathrm{t\widehat{o}p}_k$ denotes the predicted top-$k$ projects and $\mathrm{top}_k$ the real ones. Unlike the Kendall's tau coefficient, the Jaccard index does not look at the actual order but only whether a given project is included in the top-$k$ or not.
    
    
    
    \subsection{Machine Learning Predictors}\label{sec:machine_learning_baselines}
    
    
    We implemented classical supervised machine learning models using an incremental feature inclusion approach to predict the number of votes received by each project normalized by the total voter count. 
    Models were trained using a PB elections in each city and evaluated on the following PB campaign: Toulouse 2022 was used to predict Toulouse 2024; Wroclaw 2016 and 2017 were used to predict Wroclaw 2017 and 2018, respectively; and Warszawa 2021, 2022, and 2023 were used to predict Warszawa 2022, 2023, and 2024, respectively.
    
    We selected two representative architectures: a regression model (ElasticNet) and a gradient-boosted tree model (XGBoost), combined with two textual processing strategies:
    \begin{itemize}
        \item TF-IDF, from term frequency-inverse document frequency \citep{jones1972statistical}, measures how important a word is to a document by balancing how often it appears in the document against how rare it is across an entire collection of documents.
        \item BGE-M3 \citep{chen2024bge}, a multilingual encoders that transform textual data into embedding vectors, which was used to encode the full text directly. 
    \end{itemize}
    
    
    Since our PB datasets contain a limited number of projects, we performed a principal component analysis (PCA) during training for each combination of ML model/semantic embedding in order to reduce the number of features representing the textual data. To obtain robust results, we made $20$ PCA runs and perform predictions for each of them. 
    Table \ref{tab:PCA_choice_no_description} includes the median values for the PCA dimensions.

    \begin{table}[t]
    \centering
    \small
    \setlength{\tabcolsep}{4pt}

    \begin{tabular}{@{}lcc@{}}
    \toprule
    Target Instance & ElasticNet & XGBoost \\
    \midrule
    Toulouse 2024   & 2  & 4  \\
    Wroclaw 2017    & 8  & 5  \\
    Wroclaw 2018    & 2  & 20 \\
    Warszawa 2022   & 2  & 52 \\
    Warszawa 2023   & 20 & 79 \\
    Warszawa 2024   & 4  & 31 \\
    \bottomrule
   \end{tabular}
    \caption{Best-performing PCA dimensions for BGE-M3 embeddings computed from project titles only, on PB instances without project descriptions. Reported dimensions correspond to the median value over 20 runs.}
    \label{tab:PCA_choice_no_description}
    \end{table}
    
    As a benchmark, we also tested the \textsc{No-Text} configuration for both architectures where each of them was trained only on tabular data (district, category, and project cost). 
    
    All experiments were conducted using Google Colab instances. Traditional machine learning models were trained on standard runtimes equipped with an Intel(R) Xeon(R) CPU @ 2.20GHz and 12.7 GB of RAM, with per-run training times ranging from 15 to 20 minutes for Wroclaw, between 30 minutes and 2 hours for Toulouse, and between 40 minutes and 1 hour and 30 minutes for Warszawa, depending on the model and feature variant. These times refer only to the training of the ML predictors.

    \subsection{Large Language Models}\label{sec:LLMs_variants}
    \begin{figure}[t]
      \centering
        \includegraphics[width=0.48\textwidth]{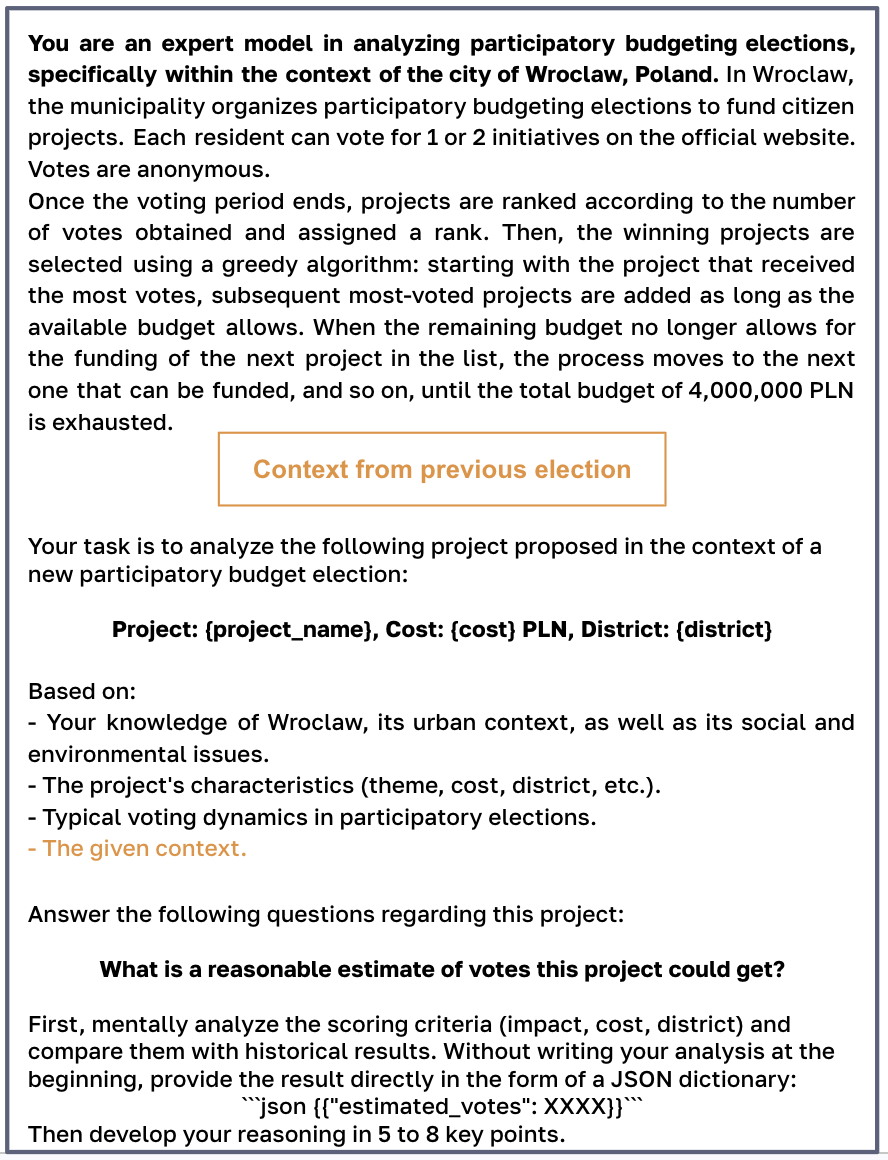}
        \caption{Prompt for the Wroclaw dataset. The \textit{context from previous election} box represents the context given by the three prompt variants considered (zero-shot, full context, RAG).}
        \label{fig:prompt}
    \end{figure}
    We design and compare three prompt variants that provide progressively richer contextual information to assess LLMs' ability to predict project support. 
    Our main hypothesis is that, when supplied with sufficient context, LLMs may capture complex multidimensional patterns that jointly reflect proposal attributes and cities' social and territorial characteristics, consistent with recent evidence from cross-modal urban data integration \citep{feng2025urbanllava}, spatio-temporal prediction \citep{li2024urbangpt}, and LLM-based urban research benchmarks \citep{feng2025citybench}.
    
    The shared prompt across all variants, illustrated in Figure~\ref{fig:prompt}, includes a brief description of the participatory budgeting process, specifies its geographical setting, and instructs the LLM to predict the total number of votes received by each project in the target PB campaigns: Toulouse 2024, Wroclaw 2017-2018, and Warszawa 2022-2024. 
    For each variant---zero-shot, full context, and retrieval-augmented generation (RAG)---we fill the context box according to the following descriptions:


    \begin{itemize}
    \item \textbf{Zero-Shot}. The LLM receives no information from previous PB campaigns, predicting the votes for each project based solely on the shared prompt. 
    \item \textbf{Full Context}. 
    The LLM is presented with the list of projects from the previous election with their title, category, cost, district of submission, and vote count. 
    \item \textbf{Retrieval-Augmented Generation} (RAG). 
    The LLM receives information on the 20\% of projects from the previous campaign that are most semantically similar to the target project (all their features: title, category, location, cost, and vote count). In addition, it also receives the three projects from the same district as the target project that are most semantically similar to it. Similar projects were computed by using the cosine similarity measure.
    \end{itemize}
    
    
    \begin{figure}
      \centering
        \includegraphics[width=0.48\textwidth]{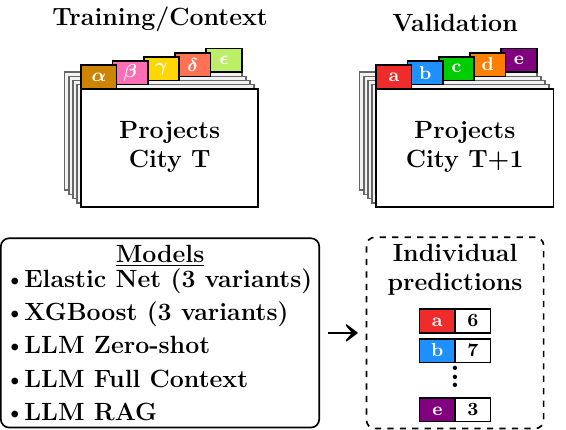}
        \caption{General workflow of our methodology to validate predictions of voting outcomes in PB elections. Projects and vote counts from a PB election are given as training data, with the objective of predicting the vote count of projects in the following election.
        }
        \label{fig:llm_voting_overview}
    \end{figure}


    

    The implementation of the three prompt variants included the \textbf{Chain-of-Thought} technique \citep{wei2022chain}, a reasoning strategy that forces the model to generate logical steps leading to a solution, as illustrated in the final part of the prompt in Figure \ref{fig:prompt}. 
    %
    %
    LLM-based predictors were accessed through external APIs and were not run locally. The LLM experiments were implemented using GPT-4o mini through the OpenAI API and Mistral Large 3 through the Mistral API. To reduce stochastic variation and facilitate reproducibility, we set the temperature to 0 for all prompts. In the zero-shot setting, 
    for a given project and PB instance, the same prompt is generated. In the full-context setting, the order in which projects are presented in the prompt is randomized using a fixed seed. In the RAG setting, retrieved projects are ranked by cosine distance, which yields a deterministic ordering for a fixed embedding space and query. 
    For all inferences, the maximum response length was constrained to 2,048 tokens. Since these settings do not guarantee strictly deterministic outputs across all providers, LLM results are averaged over multiple predictions when applicable. Despite the differences in implementation, the time required for any model (ML or LLM-based) to produce an independent prediction remained under one minute, although API rate limits could increase the wall-clock time required to collect all LLM responses. For the SBERT and BGE-M3 embedding generation, we used GPU-accelerated Google Colab instances featuring an NVIDIA A100 Tensor Core GPU.

    
    \subsection{No Prior Data Exposure} 
    
    To ensure that the language models used in this study had no prior exposure to the voting results or full project descriptions, we adapted the methodology proposed by \citet{golchin2024timetravel}. Their technique evaluates whether a model has been trained on a specific set of examples, by measuring the model’s ability to reconstruct exact or semantically similar text when presented with incomplete fragments, particularly in text completion or contextual recall.
    We selected a random sample from our dataset of projects in Toulouse 2022-2024 and Wroclaw 2016-2017, those for which project descriptions are available, focusing on their descriptions, districts, and costs, and implemented two types of test:
    
    \begin{itemize}
    \item[1.] \textbf{Project Description Completion}. We prompted the LLM to generate the full description of a project under two conditions: (a) without providing any part of the original text (only the project title), and (b) by providing the first few lines of the original description. We analyzed both lexical and semantic similarity between the real and generated descriptions. In all cases, the LLM produced generic or irrelevant content and was unable to reproduce the original description, either partially or in full.
    \item[2.] \textbf{Cost and District Retrieval:} In a second test, we provided the LLM with the project title and its full description and asked it to identify the district where the project was proposed and its estimated cost. Again, the model failed to retrieve any of this information correctly.
    \end{itemize}


    \subsection{Random Baseline}
    
    We implemented a random baseline by drawing uniformly at random a ranking of projects for each voter, a preference generation method known as the impartial culture in social choice \citep{mattei2013preflib}. For each voter, we kept the top-3 preferred projects in Toulouse, the top-2 in Wroclaw, and the top-10 in Warszawa, which in turn are aggregated to compute the predicted vote count for each project. Our results are averaged over $20$ rounds of preference generation for the instances without project descriptions and $100$ rounds for the instances with project descriptions.
    
    
    \begin{table*}[t]
\centering
\footnotesize
\setlength{\tabcolsep}{3.5pt}

    \begin{tabular}{@{}llcccccc@{}}
    \toprule
    & & \multicolumn{6}{c}{Kendall $\tau$} \\
    \midrule
    Model & Variant 
    & Toul. 24 
    & Wro.17 
    & Wro.18 
    & War.22 
    & War.23 
    & War.24 \\
    \midrule
    Random & -- 
    & 0.00 $\pm$ 0.03 
    & 0.00 $\pm$ 0.10 
    & 0.02 $\pm$ 0.08 
    & -0.01 $\pm$ 0.07
    & -0.03 $\pm$ 0.06 
    & -0.01 $\pm$ 0.05 \\
    \midrule
    \multirow{3}{*}{ElasticNet} 
    & No-Text 
    & 0.22 $\pm$ 0.02 
    & 0.33 $\pm$ 0.04 
    & 0.14 $\pm$ 0.01 
    & 0.32 $\pm$ 0.01 
    & 0.37 $\pm$ 0.03 
    & 0.25 $\pm$ 0.01 \\
    & TF-IDF 
    & 0.25 $\pm$ 0.01 
    & 0.26 $\pm$ 0.05 
    & 0.39 $\pm$ 0.02 
    & 0.37 $\pm$ 0.04 
    & 0.33 $\pm$ 0.01 
    & 0.23 $\pm$ 0.02 \\
    & BGE-M3 
    & 0.21 $\pm$ 0.06 
    & 0.16 $\pm$ 0.04 
    & 0.28 $\pm$ 0.07 
    & 0.36 $\pm$ 0.01 
    & 0.40 $\pm$ 0.01 
    & 0.30 $\pm$ 0.02 \\
    \midrule
    \multirow{3}{*}{XGBoost}    
    & No-Text 
    & 0.23 $\pm$ 0.00 
    & 0.29 $\pm$ 0.03 
    & 0.34 $\pm$ 0.00 
    & 0.37 $\pm$ 0.00 
    & 0.18 $\pm$ 0.00 
    & 0.24 $\pm$ 0.01 \\
    & TF-IDF 
    & 0.30 $\pm$ 0.01 
    & 0.44 $\pm$ 0.01 
    & 0.41 $\pm$ 0.00 
    & 0.36 $\pm$ 0.00 
    & 0.32 $\pm$ 0.00 
    & 0.24 $\pm$ 0.01 \\
    & BGE-M3 
    & 0.28 $\pm$ 0.01 
    & 0.26 $\pm$ 0.01 
    & 0.44 $\pm$ 0.01 
    & 0.40 $\pm$ 0.01 
    & 0.39 $\pm$ 0.00 
    & 0.37 $\pm$ 0.01 \\
    \midrule
    \multirow{3}{*}{\parbox{1.3cm}{GPT-4o\\mini}}   
    & Zero-shot 
    & 0.24 $\pm$ 0.02  
    & 0.36 $\pm$ 0.04
    & 0.33 $\pm$ 0.06 
    & 0.26 $\pm$ 0.03 
    & 0.31 $\pm$ 0.01 
    & 0.23 $\pm$ 0.05 \\
    & Full Context 
    & 0.22 $\pm$ 0.05
    & 0.48 $\pm$ 0.04
    & 0.43 $\pm$ 0.02
    & 0.38 $\pm$ 0.02
    & 0.44 $\pm$ 0.05
    & 0.36 $\pm$ 0.02 \\
    & RAG 
    & 0.37 $\pm$ 0.01
    & 0.34 $\pm$ 0.02
    & 0.44 $\pm$ 0.00
    & 0.44 $\pm$ 0.00
    & 0.49 $\pm$ 0.00
    & \textbf{0.53}$\pm$ 0.00 \\
    \midrule
    \multirow{3}{*}{\parbox{1.3cm}{\small{Mistral\\Large 3}}} 
    & Zero-shot 
    & 0.09 $\pm$ 0.02 
    & 0.18 $\pm$ 0.01 
    & 0.27 $\pm$ 0.05 
    & 0.40 $\pm$ 0.03 
    & 0.50 $\pm$ 0.01 
    & 0.48 $\pm$ 0.02 \\
    & Full Context 
    & 0.36 $\pm$ 0.01 
    & \textbf{0.54}$\pm$ 0.04 
    & \textbf{0.47} $\pm$ 0.04 
    & 0.47 $\pm$ 0.03 
    & 0.52 $\pm$ 0.01 
    & 0.46 $\pm$ 0.01 \\
    & RAG 
    & \textbf{0.39} $\pm$  0.00
    & 0.43 $\pm$  0.01
    & 0.44 $\pm$ 0.01 
    & \textbf{0.54} $\pm$ 0.01 
    & \textbf{0.56} $\pm$ 0.01 
    & 0.49 $\pm$ 0.01 \\
    \bottomrule
    \end{tabular}
    \caption{Kendall's $\tau$ performance of vote-based algorithmic shortlisting on the six datasets.
    Values are reported as mean $\pm$ standard deviation, computed over 20 runs for ML models and 5 predictions for LLMs. Columns correspond to Toulouse 2024, Wroclaw 2017--2018, and Warszawa 2022--2024. Results in boldface are the best model for each dataset. Results are statistically significant ($p < 0.05$), with $p$-values calculated on the rankings obtained from the average vote count of projects over the various runs of each table entry.}
    \label{tab:kendall_no_description}
    \end{table*}

    \begin{table*}[t]
\centering
\footnotesize
\setlength{\tabcolsep}{4pt}

\begin{tabular}{@{}llcccc@{}}
\toprule
& & \multicolumn{2}{c}{Kendall $\tau$}
& \multicolumn{2}{c}{Greedy Winner Position} \\
\midrule
Model & Variant & Toul. 24 & Wro. 17 & Toul. 24 & Wro. 17 \\
\midrule

Random & --
& 0.00 $\pm$ 0.00
& 0.00 $\pm$ 0.03
& 131.5 out of 183
& 17.35 out of 50 \\
\midrule

\multirow{3}{*}{GPT-4o mini}
& Zero-shot
& 0.25 $\pm$ 0.02
& 0.40 $\pm$ 0.01
& 161
& 16 \\

& Full Context
& 0.28 $\pm$ 0.02
& 0.46 $\pm$ 0.06
& 164
& 16 \\

& RAG
& 0.38 $\pm$ 0.01
& 0.22 $\pm$ 0.00
& 154
& 29 \\
\midrule

\multirow{3}{*}{Mistral Large 3}
& Zero-shot
& 0.13 $\pm$ 0.07
& 0.32 $\pm$ 0.13
& 140
& 19 \\

& Full Context
& 0.36 $\pm$ 0.01
& 0.54 $\pm$ 0.04
& 160
& 4 \\

& RAG
& 0.46 $\pm$ 0.02
& 0.33 $\pm$ 0.01
& 149
& 22 \\

\bottomrule
\end{tabular}

\caption{Performance of three prompting strategies for two LLMs on PB instances with project descriptions, for Toulouse and Wroclaw.
Metrics include Kendall's $\tau$ and the position of the last project selected under the greedy rule.
Kendall's $\tau$ is reported as mean $\pm$ standard deviation. For reference, the last project selected under the greedy rule is ranked 84th in Toulouse 2024 and 5th in Wroclaw 2017. Results are statistically significant ($p < 0.05$), with $p$-values calculated on the rankings obtained from the average vote count of projects over the various runs of each table entry.}

\label{tab:detailed_performance}
\end{table*}

    \subsection{Results}\label{sec:results}
    We evaluate our proposed pipeline for vote-based algorithmic shortlisting by comparing the predicted project support with the observed outcomes of the six PB campaigns we considered. 
    We report results for both classical ML models and LLM-based predictors in Table~\ref{tab:kendall_no_description}. 
    
    First, we observe that all our models are better predictors, and many are significantly better, than the random baseline on most measures. 
    %
    Second, the LLM-based prompting approach achieves performances comparable to, and in some cases exceeding, that of the ML models: $\tau=0.39$ in Toulouse for Mistral Large 3 RAG against $\tau=0.30$ for XGBoost in its TF-IDF variant, the best ML model for this election; or $\tau=0.56$ for Mistral Large 3 RAG against $\tau=0.40$ for ElasticNet BGE-M3 in Warszawa.
    As expected, most of the time the best results for LLM prompting are obtained by the RAG variant, as the context is carefully tailored to the project under consideration, or by the Full Context variant. 
    The latter, however, can give lower performances if the length of the context window is too large (such as in Toulouse), which can backfire and pollute the LLM reasoning \citep{liu2024lost}. 
    We still observe that the LLM predictions are surprisingly accurate,
     given that LLMs are made for predicting text and not scalars such as the vote count (see the discussion in Section~\ref{sec:conclusions_discussions} on different types of prompts), unlike the classical ML methods we implemented.
    
    %
    
\begin{figure}[t]
        \centering
        \includegraphics[width=0.48\textwidth]{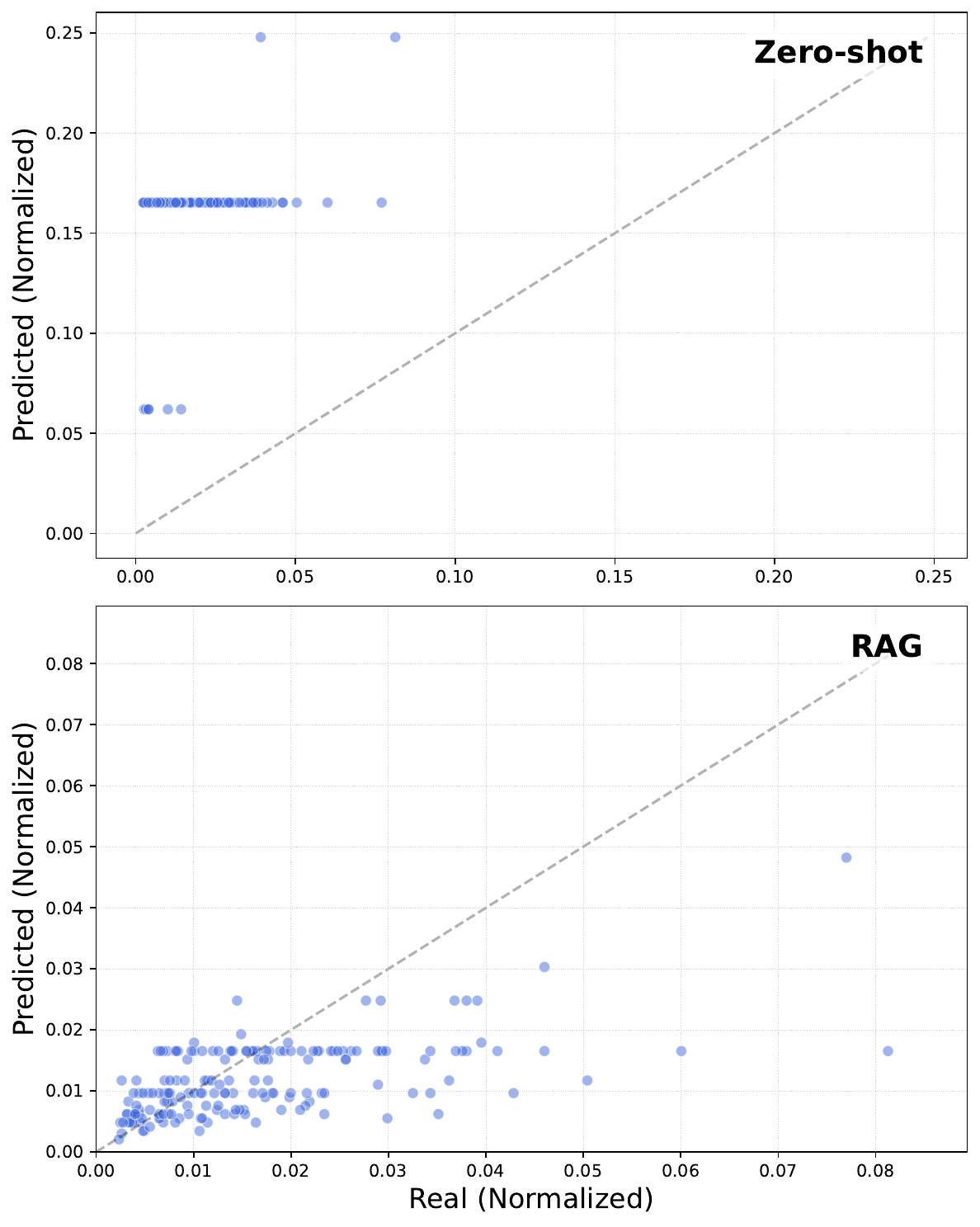}
        \caption{Predicted vs. real vote counts, normalized by number of voters for Mistral Large 3 on the Toulouse dataset. The figures compare zero-shot (top) and RAG (bottom) prompts. The dashed line is the identity line ($y=x$), indicating perfect prediction.
        }
        \label{fig:zeroshotRAG}
    \end{figure}

    Digging further to analyze the LLM predictions, Figure \ref{fig:zeroshotRAG} shows the predicted and real vote count of Mistral, respectively for zero-shot and RAG prompts.
    Without the context, the zero-shot prompt variant divides all projects into classes and predicts the same number of votes for each project within the same class, with a quite inaccurate scale for the number of votes. 
    This shows that only prior knowledge about the city is not sufficient to retrieve the voters' opinions. Moreover, it showcases that LLMs have an impressive capacity of aligning their predictions to the problem at hand when provided relevant context. 
    
    
    \subsection{Adding Project Descriptions}
    
    
    For two datasets, Toulouse 2022--2024 and Wroclaw 2016--2017, we were able to complement our data project textual descriptions. Table~\ref{tab:detailed_performance} reports the results on these two datasets. Given the additional textual component, we only run the LLM-based pipelines. In addition to the Kendall-Tau distance, we also compute the lowest position of an actual winning project in the predicted ranking, to obtain an estimation of how coarse a shortlisting can be to include all the winning projects.
    
    First, Comparing Tables~4 and~5, we observe that adding project descriptions improves or maintains performance across all settings, with the exception of Mistral in Wroclaw. For Toulouse, the additional textual information improves all models, with Mistral Large 3 RAG increasing from $\tau = 0.39$ without descriptions to $\tau = 0.46$ with descriptions. For Wroclaw, however, adding project descriptions can be detrimental: the Mistral RAG score decreases from 0.43 in Table~4 to 0.33 in the augmented setting. Second, the random baseline suggests that the Toulouse PB election results are easier to predict than the Wroclaw one (lowest position of a greedy winner at 131 out of 183 in Toulouse against 17 out of 50 in Wroclaw). 
    This can be explained by the fact that in Wroclaw the budget is smaller and so the number of accepted projects is lower (72 accepted projects out of 183 in Toulouse-2024, 5 accepted projects out of 50 in Wroclaw-2017). Moreover, the high variability of the results in-city and between-city show that a shortlist that contains all real winning projects might actually be of size comparable to that of all projects, limiting the gains of algorithmic shortlisting.
    \section{Conclusion and Discussion}\label{sec:conclusions_discussions}
    
    We introduced an innovative methodology for algorithmic shortlisting in PB processes, comparing traditional ML methods with three LLM prompts on multi-year PB datasets from three cities.
    %
    Our results show that LLMs can be prompted effectively to obtain predictions of the ranking of most-voted projects, with context-aware prompting strategies obtaining competitive and stable performances. Our findings show that LLMs can be a viable tool for PB organisers, as their natural language interface lowers the technical barriers to deploying algorithmic shortlisting tools.
    Their main limitation occurs when coupled with a voting rule such as the widely used greedy rule, where we highlighted a limited applicability to find a shortlist guaranteeing that all winning projects are present. 
    Still, our tools can support innovations in digital democracy by helping PB organizers manage large numbers of proposals, as our pipelines can recover a substantial share of the top projects in the real rankings across the considered PB instances.
    Importantly, our proposed methodology has low data-collection barriers, and LLM-based implementations require fewer AI skills than traditional predictive pipelines, encouraging adoption by PB actors.
    
    \noindent\textbf{Demographics-based and voter-level predictions}.
    Voting rules for participatory budgeting such as the methods of equal shares \citep{PierczynskiSP21} require the full information on individual ballots to be computed (basically, which voters approved which project). 
    Previous work by \citet{yang2024llm} and by \citet{gudino2024large} attempted to obtain voter-level predictions by constructing LLM personas from the demographics of the participants---the first paper using data from a PB experiment, the second using pairwise comparisons over political proposals.
    Both papers indicate that accuracy for voter-level prediction is low (still,  \citet{gudino2024large} shows accuracies between 68\% and 77\% for individual choices, with slightly lower performance when comparing proposals from the same political candidate).
    Moreover, \citet{yang2024llm} showed important order-effects in PB elections, due to the ordering of presentation of projects on the ballots.
    However, at the aggregate level, LLM predictions recover the ground truth more reliably.
    %
    A direct comparison between \citet{yang2024llm}'s results and our approach is not possible, as their data is a single-year instance.
    However, our results on the ordinal ranking are comparable to their aggregated findings (see Figure 4 in \citet{yang2024llm}), 
    where GPT-4o mini achieved a Kendall-$\tau$ of $0.41$ against a human baseline, while we obtain results between $0.29$ and $0.44$ with RAG prompts, and $0.35$ and $0.38$ with Full Context prompts.
    This suggests that demographic data may be less effective than project descriptions and historical election data in predicting project performance, while being publicly accessible (a finding also confirmed by \citet{gudino2024large}'s ablation studies).

    \noindent\textbf{Cost bias}. 
    In most real-world instances the ratio between the vote count of a project and the total number of voters is close to the ratio between its cost and the total budget \citep{schmidt2024cost}. 
    This means that running a greedy mechanism over projects ranked by their vote count produces biased results toward higher-cost projects. 
    %
    We have preliminary evidence that helping the ML methods to focus on cost can improve their performance, and future work could test whether this is also true of LLM prompting (e.g., with a cost-oriented RAG).
    
    \noindent\textbf{Type of prompt}. Using the terminology of \citet{licht2025measuring}, our prompts implement \emph{pointwise scoring} (scalar prediction) and compare three strategies: zero-shot, few-shot, and many-shot in-context learning. Since LLMs are often more effective at \emph{pairwise scoring}, future work could compare our approach to prompts that ask which of two projects is more likely to receive more votes, avoiding cardinal predictions at the cost of substantially more LLM queries to recover a full ranking. 
    Alternatively, posterior pairwise scoring among the top 30\% predicted most-voted projects with our methods, could be tested to recover the ranking of the highest scored projects.

    \noindent\textbf{Limitations}.
    GPT-4o mini is a commercial LLM and may not be accessible to municipalities running PB campaigns. Open-source LLMs such as Mistral Large 3 are alternative options, and complementary experiments show they have similar and sometimes better performances.
    However, our results show that LLMs cannot be used as out-of-the-box shortlisting tools by PB organizers but require some technical skills to set up, e.g., a RAG. Future work can experiment with more easily accessible prompting strategies building on the Full Context prompting results.
    
    Our approach, like any shortlisting method based on historical data, assumes a certain degree of stability across consecutive PB processes. Social, economic, or political changes may 
    alter citizens' preferences, reducing the predictive capacity of any model, whether based on classical ML or LLMs. 
    Such stability assumption is more plausible in online PB processes in the Global North, where the stakes are often relatively low, as in the three European cities we considered.
    Thus, validation across different institutional and geographical contexts is one of our primary directions for future work.
    
    
    We did not assess the robustness of our methods to strategic user behavior, such as using specific language in project descriptions or voting tactically based on visible vote counts. 
    Overreliance on algorithmic predictors at the proposal stage could even encourage similar examples of strategic behavior by proposers, but also reduce their predictive power over time or undermine the expected diversity of projects. 
    
    Finally, our approach to shortlisting optimizes exclusively with respect to predicted electoral support, reinforcing a status-quo bias. An evident example is that projects that have already been funded in a previous campaign would score high, but should clearly not be shortlisted.
    However, algorithmic shortlisting may and should 
    incorporate additional normative criteria, such as representational fairness or diversity preservation.
    Our predictions can only be used as one of the factors in a multi-objective algorithmic shortlisting mechanism, generating an initial ranking for human review as part of an AI-assisted democratic process.

    

    
    %
    %
    %

    \section*{Ethical Statement}
    \medskip
    \noindent \textbf{Data ownership}. 
    The data collected for this paper is publicly available on Pabulib.org. The datasets including project textual information were obtained by combining voting data from the mentioned website with project descriptions scraped from the public webpages of the cities of Wroclaw and Toulouse.
    
    \medskip
    
    \noindent\textbf{Potential risks}.
    The use of LLMs in democratic applications raises ethical concerns, with examples such as pre-existing and documented biases of LLMs, the possibility of result manipulation, or the replacement of actual voters by AI agents at the voting phase. Our results are to be considered as an exploratory proof-of-concept which requires in-depth further scrutiny before being deployed or tested with human subjects.
    The use of black box ML algorithms, LLMs, or even interpretable methods that are still not accessible to the ordinary citizen is debatable. Still, if its use is accompanied by proper detailed documentation, our approach defines a shortlisting procedure that may be more transparent than some current practices, which often involve committees of volunteers selecting projects on loosely specified objectives.
    
    \section*{Acknowledgments}
    
    \noindent This work was supported by the ANR LabEx CIMI (grant ANR-11-LABX-0040) within the French State Programme “Investissements d’Avenir.” 
    
    \noindent Funded by the European Union. Views and opinions expressed are, however, those of the author(s) only and do not necessarily reflect those of the European Union or the European Research Council Executive Agency. Neither the European Union nor the granting authority can be held responsible for them. This work is supported by ERC grant 101166894 “Advancing Digital Democratic Innovation” (ADDI). This project was supported by the European Union LearnData, GA no. 101086712 a.k.a. 101086712-LearnDataHORIZON-WIDERA–2022-TALENTS–01 (\url{https://cordis.europa.eu/project/id/101086712}), IAST funding from the French National Research Agency (ANR) under grant ANR–17-EURE–0010 (Investissements d'Avenir program), and the European Lighthouse of AI for Sustainability [grant number 101120237-HORIZON-CL4–2022-HUMAN–02].
    \par\noindent
    \makebox[\columnwidth][c]{%
      \includegraphics[width=0.40\columnwidth]{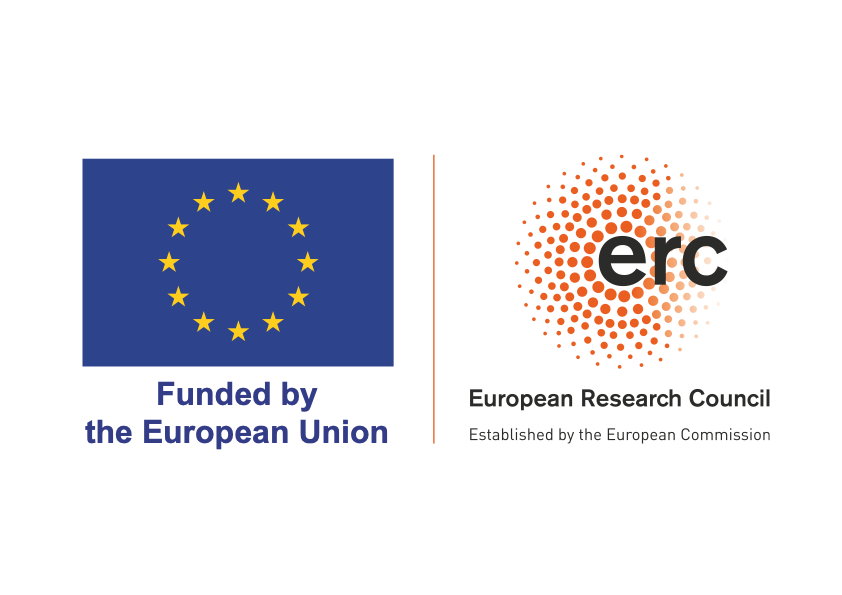}%
    }




%
%
%

\bibliographystyle{apalike}
\bibliography{aaai2026}

\appendix

\section{Datasets Description}\label{app:datasets}
The datasets used for our study are constructed combining data publicly available on Pabulib.org for the three cities considered in this paper (project titles, categories, locations, estimated costs, and vote counts). For four instances, we additionally collected project textual descriptions from the webpages of the municipality of Toulouse\footnote{\url{https://jeparticipe.metropole.toulouse.fr/processes/mes-idees-pour-mon-quartier} \\  \url{https://jeparticipe.metropole.toulouse.fr/processes/bp2023}} and Wroclaw,\footnote{\url{https://www.wroclaw.pl/rozmawia/wroclawski-budzet-obywatelski-edycja-2016} \\ \url{https://www.wroclaw.pl/rozmawia/wroclawski-budzet-obywatelski-edycja-2017}} accessed in August 2025. \\

We created one project dataset per city: Toulouse, France; Wroclaw, Poland; and Warszawa, Poland. 
Each city dataset contains the PB instances considered for that city: Toulouse 2022 and 2024; Wroclaw 2016, 2017, and 2018; and Warszawa 2021, 2022, 2023, and 2024. 
For all cities and PB campaigns, we consolidated datasets containing the following attributes: Project ID, Title, Category, Estimated Cost, District, and Vote Count. 
For Toulouse 2022--2024 and Wroclaw 2016--2017, Textual Description is included as an additional column in the corresponding city dataset. For Warszawa, district information was available in the raw data but was not used either in the prompts or in the ML models.

\medskip

Tables~\ref{tab:participation_summary} and \ref{tab:project_examples_tls} provide examples of projects from the city of Toulouse in 2022 and from Wroclaw in 2017.

\begin{table}[h!]
  \centering
  \caption{Examples of lines in the Toulouse 2022 dataset.}
  \label{tab:project_examples_tls}
  \begin{small}
    \begin{sc}
      \begin{tabularx}{\textwidth}{l p{2.5cm} X l r c r}
        \toprule
        ID & Name & Description & Category & Cost & Dist. & Votes \\
        \midrule
        136 & Bike lane on Saint-Exupéry Avenue & When you live near Place de l'Ormeau, cycling into town is dangerous. Wouldn't it be possible to consider creating a proper bike lane there? ... & Eco-mobility & 200 000 & 1 & 492 \\
        \addlinespace
        17 & Fighting mosquitoes with bats & Bats could eat up to the equivalent of 3000 insects per night (including mosquitoes). ... & Nature in the City & 1 200 & 2 & 155 \\
        \bottomrule
      \end{tabularx}
    \end{sc}
  \end{small}
  \end{table}
  
  \begin{table}[h!]
  \centering
  \caption{Examples of lines in the Wroclaw 2017 dataset.}
  \label{tab:project_examples_wro}
  \begin{small}
    \begin{sc}
      \begin{tabularx}{\textwidth}{l p{2.5cm} X l r c r}
        \toprule
        ID & Name & Description & Category & Cost & Dist. & Votes \\
        \midrule
        201 & Ping-pong tables and volleyball courts & Movement is health. In Wroclaw, there is a high demand for places where one can actively spend time outdoors. ... & Sport & 1 000 000 & 1 & 3 974 \\
        \addlinespace
        694 & "Green stops" in the city center & The project proposes the creation of green (bus) stops at selected locations in downtown Wroclaw by planting climbing plants... & Greenery & 200 000 & 1 & 943 \\
        \bottomrule
      \end{tabularx}
    \end{sc}
  \end{small}
\end{table}

The Project ID serves as a unique identifier within each city’s participatory budgeting cycle. The Name represents the title written by citizens during the proposal phase. For the instances with project descriptions, the Textual Description represents the detailed explanation associated with each project. The descriptions have a mean length of 140 words in Wroclaw and 150 words in Toulouse (approximately 1000 characters). 
Projects are also classified into categories (e.g., Urban Nature, Eco-mobility) and assigned an estimated cost by the election organizer. The district attribute indicates the geographical scope of the project, with 20 districts in Toulouse, 5 districts in Wroclaw, and 18 districts in Warszawa. 
Finally, the vote count represents the primary target variable, indicating the total number of citizen votes received by each project. \\

The three consolidated city-level datasets (Toulouse, Wroclaw, and Warszawa) are available in the supplementary material of this submission.

\section{Naive Shortlisting Prompt Templates}

We evaluated two direct shortlisting prompts: a zero-shot prompt and a many-shot prompt. In both settings, the model is asked to select a fixed number of projects expected to be the most popular in the target PB campaign. This number is computed from a shortlist percentage $k \in \{10\%,20\%,30\%,50\%\}$ and the total number of projects in the corresponding election. 

\medskip

We denote the resulting number of projects by $n_k$. 
The prompt then asks the model to return exactly $n_k$ project identifiers, sorted from most likely to least likely in expected number of votes.The model is instructed to first return a valid JSON dictionary containing only project identifiers, and then provide a short explanation in 5 to 8 key points. 

\begin{figure*}[t]
    \centering

    \begin{minipage}{0.48\textwidth}
        \centering
        \includegraphics[width=\linewidth]{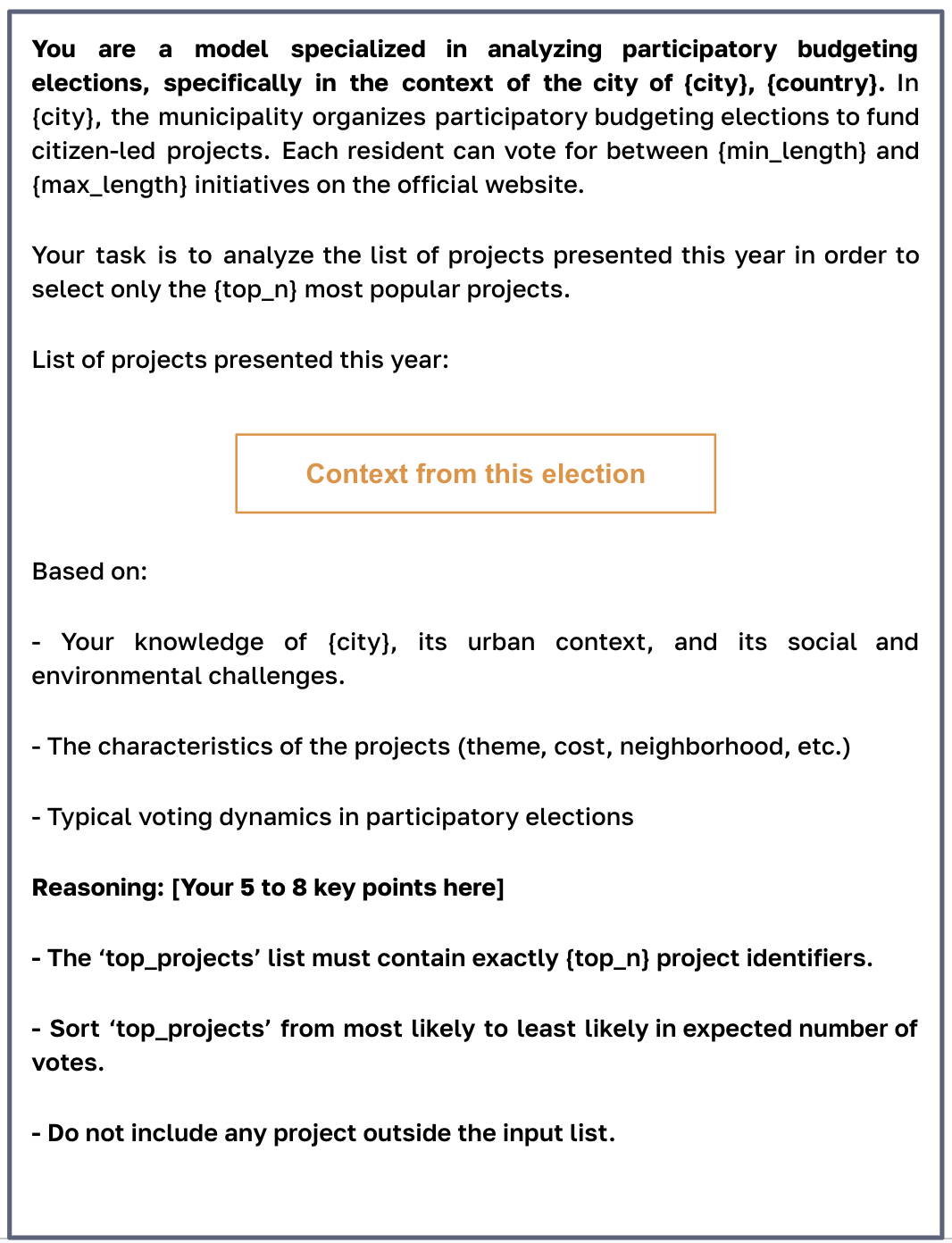}\\
        \small (a) Zero-shot shortlisting prompt
    \end{minipage}
    \hfill
    \begin{minipage}{0.48\textwidth}
        \centering
        \includegraphics[width=\linewidth]{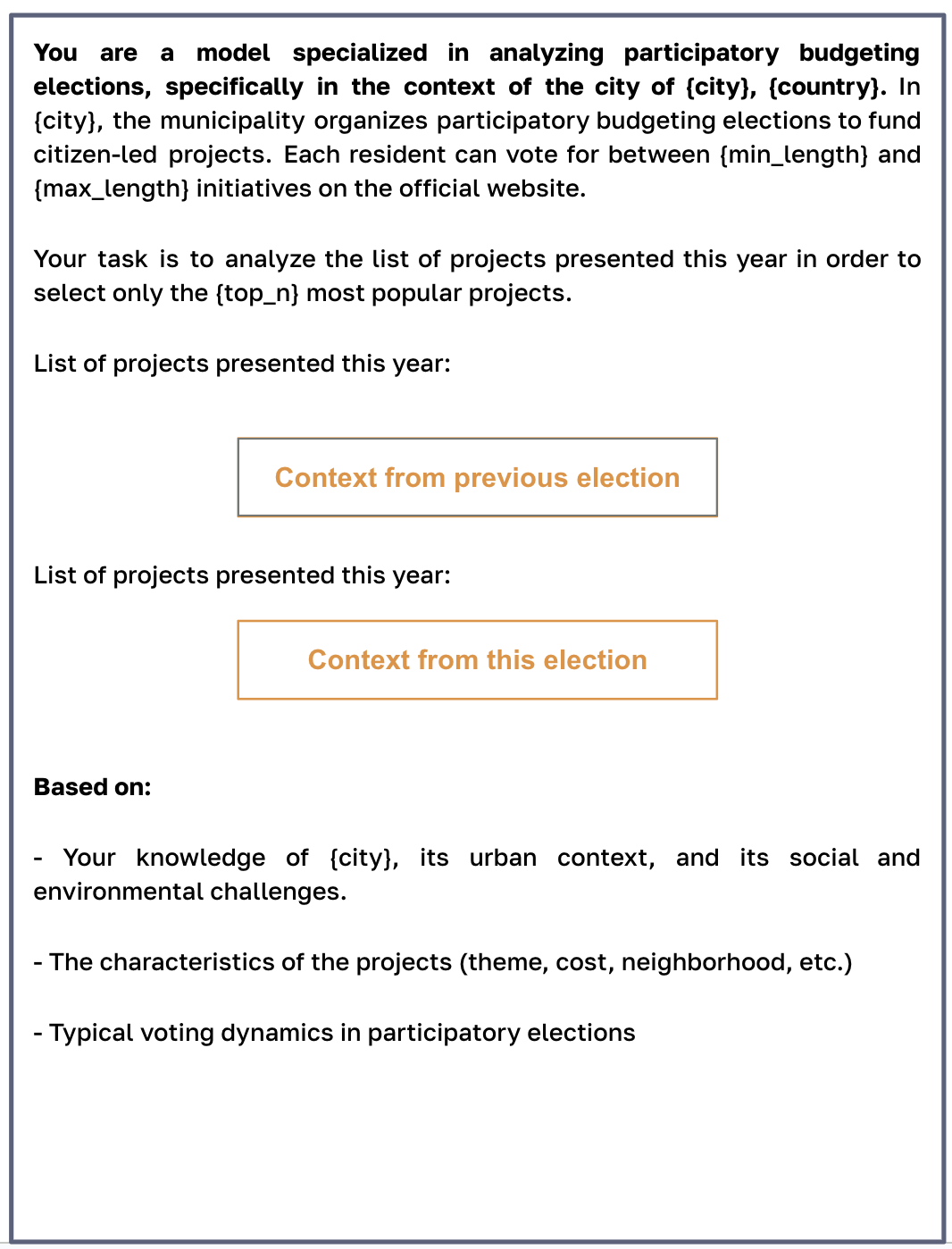}\\
        \small (b) Many-shot shortlisting prompt
    \end{minipage}

    \caption{Prompt templates used for the direct shortlisting task. The zero-shot prompt receives only the list of projects from the target PB campaign, while the many-shot prompt additionally receives the projects and observed vote counts from the previous campaign.}
    \label{fig:shortlisting_prompts}
\end{figure*}

Figure~\ref{fig:shortlisting_prompts} shows the prompt templates used in both settings.

\paragraph{Zero-shot shortlisting prompt.}
In the zero-shot setting, the model receives the list of projects presented in the target year:
\[
\texttt{\{project\_list\_shuffled\}}.
\]
The order of this list is randomized using a fixed seed. 
Based on the city context, project characteristics, and typical voting dynamics in PB elections, the model is asked to select exactly $n_k$ projects expected to be the most popular.

\paragraph{Many-shot shortlisting prompt.}
In the many-shot setting, the model receives the list of projects from the previous PB campaign, together with their observed vote counts:
\[
\texttt{\{last\_year\_project\_list\_shuffled\}},
\]
as well as the list of projects from the target year:
\[
\texttt{\{project\_list\_shuffled\}}.
\]
The order of both lists is randomized using a fixed seed. 
The model is asked to use the previous year's results and the target year's project list to select exactly $n_k$ projects expected to be the most popular in the target campaign.

\section{Data Preparation for Machine Learning Predictors}

For both machine learning models used in this study (ElasticNet and XGBoost) we transformed the raw project data into a structured feature set. 
First, 
we processed the categories and districts features using one-hot encoding and we applied a logarithmic transformation to the project cost to better align its scale with the distribution of the target variable. 
The target variable was defined as the total vote count obtained by each project, normalized by the city's total voter turnout for the respective year. 
These basic features were the only ones available to the \textsc{No-Text} variant.

To evaluate the impact of the projects' semantic content, we extended the base feature set using text representation strategies applied to project titles. 
For the instances with project descriptions (Toulouse 2022--2024 and Wroclaw 2016--2017), we additionally applied these strategies to the textual descriptions.

\begin{enumerate}
\item TF-IDF: We calculated term importance based on statistical frequency. 
The TF-IDF vectorizer was configured to convert text to lowercase and extract both unigrams and bigrams ($n$-gram range from 1 to 2). 
We limited the vocabulary to the 20,000 most frequent tokens to maintain a manageable feature space. 
To mitigate the influence of repeated terms within a single project text, we applied sublinear term frequency scaling ($1 + \log(\text{tf})$). 
For the Toulouse dataset, we also enabled Unicode accent stripping to ensure consistency across French character variants. 
All resulting features were stored as 32-bit floating-point numbers to optimize computational efficiency.

\item Sentence Embeddings: We used pre-trained language models to map text into a continuous vector space. 
Specifically, we employed SBERT \citep{reimers-2019-sentence-bert} and BGE-M3 \citep{chen2024bge}. 
Since the Toulouse, Wroclaw, and Warszawa datasets are in French and Polish, the use of multilingual models was important for capturing the semantic meaning of the original proposals without relying on machine translations. 
\end{enumerate}

We included SBERT using a multilingual BERT architecture to maintain a comparative baseline close to the original BERT framework, adapted to the nature of the text of PB campaigns. 
Second, we incorporated BGE-M3 as a modern LLM-based embedding alternative optimized for multilingual tasks. 
We also evaluated OpenAI's text-embedding-3-large model; however, it was discarded after yielding inferior results in preliminary tests compared to the selected models.

\end{document}